\documentstyle[12pt]{article}    

\begin{document}



\centerline{\large\bf On the Hadronic Beam Model} 
\centerline{\large\bf for Gamma-ray Production in Blazars}
\vspace{1truecm}
\vspace{1truecm}
\centerline{J.H. Beall$^{1,2,3}$ \& W. Bednarek$^{4}$}

\vspace{1truecm}


\centerline{$^1$E.O. Hulburt Center for Space Research,} 
\centerline{Naval Research Laboratory, Washington, DC 20375} 
\centerline{$^2$CSI/CEOSR, George Mason University, Fairfax, VA 22030} 
\centerline{$^3$St. John's College, Annapolis, MD 21404, USA}
\centerline{$^4$Department of Experimental Physics, University of \L
\'od\'z,} 
\centerline{ul. Pomorska 149/153, PL 90-236 \L \'od\'z, Poland }

\vspace{2truecm}

\begin{abstract}

We consider, herein, a model for $\gamma$-ray production in blazars in which
a relativistic, highly-collimated electron-proton beam interacts with a dense,
compact cloud as the jet propagates through the broad and perhaps narrow
line regions (BLR and NLR) of active galactic nuclei (AGN).  During the
propagation of the beam through the cloud, the process of excitation of
plasma waves becomes an important energy loss mechanism, especially for
mildly relativistic proton beams. We compute the expected spectra of
$\gamma$-rays from the decay of neutral pions produced in hadronic
collisions of the beam with the cloud, taking into account collisionless
losses of the electron-proton beam. This model may explain the X-ray and TeV
$\gamma$-ray (both low and high emission states) of Mrk 421 as a result of
synchrotron emission of secondary pairs from the decay of charged pions and
$\gamma$-ray emission from the decay of neutral pions for the plausible
cloud parameters. However clouds can not be too hot and too dense. Otherwise
the TeV $\gamma$-rays can be attenuated by the bremsstrahlung radiation in
the cloud and the secondary pairs are not able to efficiently produce
synchrotron flares because of the dominant role of inverse Compton
scattering.

The non-variable $\gamma$-ray emission observed from Mrk 421 in the EGRET
energy range cannot be described by the $\gamma$-rays from decay of neutral
pions provided that the spectrum of protons in the beam is well described by
a simple power law. These $\gamma$-rays might only be produced by secondary
pairs scattering the soft non-variable X-rays which might originate in the
inner part of the accretion disk.

\end{abstract}

\vspace{0.4truecm}

\noindent
{\it Subject headings:} galaxies: active - galaxies: jets - BL Lacertae objects: 
individual (Mrk 421) - gamma-rays: theory
\vspace{0.4truecm}

\newpage

\section{Introduction}

Many blazar-type active galaxies have recently been detected in MeV - GeV
$\gamma$-rays by detectors on the board of Compton GRO (von Montigny et
al.~1995). Mrk 421, Mrk 501, and 1E2344 are observed in the TeV
$\gamma$-rays (Punch et al.~1992, Quinn et al.~1996, Catanese et al.~1997a). 
The $\gamma$-ray emission of blazars is highly variable on different time
scales (from months and weeks up to a fraction of an hour, e.g. Gaidos et
al. 1996, Mattox et al. 1997). The spectra of FSRQ blazars are well
described by a power-law with a spectral break from the lower energy part
observed in a few sources (McNaron-Brawn et al. 1995, Sch\"onfelder et al.
1996), and possibly a cut-off at higher energies (Pohl et al. 1997). The
spectra of two BL Lacs, Mrk 421 and Mrk 501, extend up to at least $\sim 10$
TeV (Krennrich et al. 1997, Aharonian et al. 1997). There is also weak
evidence of emission of $\sim 50$ TeV photons from some nearby BL Lacs
(Mayer \& Westerhoff 1996).

These observations are usually interpreted in terms of the inverse
Compton scattering model.  This model has been discussed previously in many
different geometrical scenarios (see for example reviews by Dermer \&
Schlickeiser 1992, Sikora 1994, Dermer \& Gehrels 1995, Schlickeiser 1996,
Bednarek 1998). The $\gamma$-rays can be also produced in collisions of very
high-energy hadrons with soft radiation (e.g. Sikora et al. 1987). This
mechanism is discussed in the context of $\gamma$-ray production in blazars
by e.g.  Mannheim \& Biermann (1992), Coppi, Kartje, K\"onigl~(1993),
Protheroe (1997), Bednarek \& Protheroe (1997).

The interaction of hadronic jets with background matter in AGNs as a
possible source of $\gamma$-rays has been problematic because of the
difficulty of finding a dense enough target. In principle, such a target
might be created either by the matter of a thin accretion disk (Nellen et
al. 1993) or by the matter in a corona of a thick accretion disk (Bednarek
1993). The $\gamma$-ray emission observed from blazars might originate in
hadronic collisions of a highly collimated proton beam with clouds entering
the jet.  This model, recently proposed by Dar \& Laor (1997), is similar to
the scenario developed by Rose et al.~(1984, 1987).

In the broad line region (BLR), the typical dimensions of clouds are of
order $\sim 10^{12-13}$ cm with cloud densities $\sim 10^{10-12}$ cm$^{-3}$. 
For these parameters, the hadronic collisions occur frequently enough to
produce the observed fluxes of $\gamma$-rays.  

However, collisionless processes principally driven by the two-stream
instability (Scott, et al.~1980, Rose et al.~1984, 1987) can be the dominant
energy loss mechanism for such jets in parameter ranges associated with the
BLR and NLR (Beall 1990). The computations by Dar and Laor do not take into
account these collisionless energy losses as the jet propagates through the
dense, hot medium of the cloud.

Where collisionless  energy losses are significant, propagation lengths are
markedly shortened.  This is especially true for electron-proton beams with
relatively low Lorentz factors (see Rose et al. 1984, and Beall 1990). 
These effects must be taken into account when calculating the $\gamma$-ray
emission from the jet.

The radiation emitted by a relativistic electron-proton beam interacting
with an ambient medium such as a broad line cloud has been previously
discussed in general terms by Rose et al.~(1987) and Beall et al. (1987). 
In this paper the predictions of the model wherein an electron-proton beam
interacts with an interstellar cloud are compared with the results of
$\gamma$-ray observations of Mrk 421 in a high and low states.  We extend
the results of the recent discussion of this model by Dar \& Laor (1997) to
take into account the influence of collisionless losses on the
electron-proton jet's $\gamma$-rays spectra.

\section{Propagation of Relativistic Electron-Proton Beams Through 
Interstellar Clouds} 

Rose et al.~(1984, 1987) first suggested that the interaction of
relativistic particles with dense, interstellar clouds could account for the
variability and flux of the hard x-ray and $\gamma$-ray sources in active
galaxies, and discussed in detail the mechanisms of energy loss for a
relativistic, low-density beam of electrons, electrons and positrons, or
electrons and protons as it interacts with clouds in the BLR and NLR of AGN.

As the beam of relativistic  particles deposits energy  in the ambient
medium via the generation of electrostatic plasma waves, a number of
important physical processes are operant.  The material in the jet cone or
cylinder suffers periodic acceleration as the two-stream instability
generates regions of high electric field intensity which then further
``sweep out" electrons (and eventually background atoms) from the region
where the high electric fields are generated.  These ``cavitons" are
low-density, microscopic structures that have a net motion with respect to
the ambient medium. During the time when they form, evolve, and then
collapse (much like a wave breaking on a shore), they transfer momentum to
the ambient medium in the direction of the jet's motion.

In addition to the transfer of momentum to the ambient medium by the net
motion of the electric fields in the cavitons, a high-frequency component of
the caviton electric field interacts with electrons in the ambient medium in
a manner that evolves the Maxwell-Boltzmann distribution of the gas,
producing a high-energy tail.  This high-energy tail is a remarkably
efficient mechanism for ionizing the gas in the Broad and Narrow Line
Regions of AGN (Beall and Guillory, 1996).  In addition, the high-energy tail on the Maxwell-Boltzmann
distribution can decrease the growth rate of the parameteric (Oscillating
two-stream) instability (Freund, Smith, Papadopoulos, and Palmadesso 1981)
and thus effects the heating rate of the beam upon the plasma, and the
cooling rate via inelastic processes and radiation transport.

In order to calculate the propagation length of the electron-proton jet
described above, we model the interaction of the relativistic jet with the
ambient medium through which it propagates by means of a set of coupled,
partial differential equations which describe the growth, saturation, and
decay of the three wave modes likely to be produced by the jet-medium
interaction.  First, two-stream instability produces a plasma wave, $W_1$,
called the resonant wave, which grows initially at a rate
$\Gamma_1=(\sqrt{3}/2\gamma )(n_{b}/2n_{p})^{1/3}\omega_{p}$, where $\gamma$
is the Lorentz factor of the beam, $n_{b}$ and $n_{p}$ are the beam and cloud
number densities, respectively, and $\omega_{p}$ is the plasma frequency, as
described more fully in Rose et al.~(1984).  Secondly, a parametric
instability fed by this wave generates high-frequency components to the
electrostatic waves in the plasma.  These high-frequency components we
designate as $W_2$.  Finally, the interaction of these waves with one
another and with the ions in the ambient medium generate ion-acoustic waves,
which we designate as $W_s$.  Such a treatment of the process considered
herein does not of course provide a self-consistent calculation of the
deposition of the beam energy into the plasma.  It does, however, provide a
reasonable estimate of the magnitude of the effects of the interaction of
the relativistic jet with the ambient medium.  A more detailed discussion of
the considerations leading to these rate equations can be found in Scott et
al.(1981), Rose et al.(1984, 1987), and Beall et al.(1986, 1990).  The
results of these estimates have been substantially confirmed by a
Particle-in-Cell Code (PIC-Code) calculation (Rose et al.~ 1998, Beall et
al.~1997).  The introduction of a weak, longitudinal magnetic field tends to
stabilize the beam and increase its propagation length somewhat.

The time-dependent values of the normalized wave energy densities for the
two-stream instability, $W_1$, the oscillating two-stream instability,
$W_2$, and the ion-acoustic waves, $W_s$, and the shortest time scale for
the growth of the instabilities, t, determine the rate at which energy is
drawn from the relativistic beam into the ambient medium.  We note that
$\tau=1/\Gamma_1$, where $\Gamma_1$ is the initial growth rate of the two-stream
instability wave energies.  These solutions represent a  spatial average of the  energy density for
the waves in the plasma.

As a beam excites waves in traversing a background plasma, it loses
energy and $\gamma$ decreases.  For an electron-proton beam the principal
collisionless interaction is between the beam electrons and background
plasma.  Consequently, the beam electrons will tend to slow down with
respect to the beam protons.  If the beam-plasma interaction is not very
strong, the beam protons will drag the electrons along with them.

For a relativistic beam $v_{b}=c(1-v^{2}/c^{2)^{-1/2}}$
and the energy  loss  through  a  distance $\Delta l$ can be estimated as
follows.  Let

\begin{eqnarray}
v_{b}n_{b}m^{'}c^{2}(d\gamma /dx)\Delta \simeq -(d(\alpha e_{1})/dt
\end{eqnarray}

\noindent where $m' = m$ for an electron beam, $m' = m$ for an
electron-positron beam, $m' = mp + m$ for an electron-proton beam,
$\epsilon_1$ is the energy density of the resonant waves, and $\alpha$ is a
factor ($\geq$1) that corrects for the simultaneous transfer of resonant wave
energy into nonresonant wave energy $\epsilon_2$ and ion-acoustic wave
energy $\epsilon_s$.  From equation 1, we find

\begin{eqnarray}
dE/dx=-(1/n_{b}v_{b})(d\alpha _{1}/dt), and \\ 
\int d\gamma =-\int [d(\alpha \epsilon _{1})/dt]/(v_{b}n_{b}m^{'}c^{2})
\end{eqnarray} 

\noindent 
This equation can be compared directly with the
equations for the individual particle energy loss mechanisms (Beall 1990). 
We note that $\alpha\epsilon_1 = \alpha W_1/nkT$, and that the $dt$ is the
cycle time between periods in the oscillatory solutions to the equations,
or, for the steady state solutions, roughly $1/\Gamma_1$.  The two time
intervals are approximately equal.

Given these considerations, the propagation length Lp [i.e., the distance
over which \( \gamma =(1-v^{2}/c^{2})^{-1/2} \) decreases by a factor of
$\sim 2$] becomes

\begin{eqnarray}
L_{p}=(\gamma /2)v_{b}n_{b}m^{'}c^{2}/<d\alpha e _{1}/dt>
\end{eqnarray}

\noindent
where $<d\alpha e _{1}/dt>$ is the time  average  rate  of excitation of
wave energy density, can be obtained from the solutions described below.

Figure 1 shows the dependence of propagation length, $L_p$, vs. $\gamma_b$
for a range of parameters appropriate for the BLR of AGN.

\section{Gamma-rays from Interaction of Electron-Proton Beams with Dense Clouds}

It is plausible that highly collimated, relativistic proton beams can be
accelerated in the strong electric fields created by a black hole or an
accretion disk rotating in the perpendicular magnetic field (see
e.g. Blandford \& Znajek 1977, Lovelace 1976, Blandford 1976), or during
magnetic reconnection occurring in the jet or on the surface of an accretion
disk (e.g. Romanova \& Lovelace 1992, Haswell et al. 1992).  In the second
scenario both electron and proton beams can be formed in multiple
reconnection events occurring at this same time on the inner part of an
accretion disk.  Unfortunately, there is at present no theory which is able
to predict the spectrum of particles accelerated via such  magnetic
reconnections. 

The spectrum of protons injected from the single reconnection
region might be close to monoenergetic, if all particles pass through this
same electric field potential, or it might resemble a power law if the
escape of particles from the reconnection region is stochastic (Colgate
1995). Even if the proton spectrum from a single reconnection region is
monoenergetic, the total spectrum summed up over many reconnection regions
might resemble a power law. Therefore it seems reasonable to
consider a power law spectrum of the proton beam,  as Dar
\& Laor (1997) have done. The electron-proton (e-p) beams, propagating in the
surroundings of active galaxy, suffers energy losses via collisionless
excitation of plasma waves (op.cit.) and through hadronic collisions with matter. In
hadronic collisions, neutral and changed pions are created. These decay to
electrons and positrons, $\gamma$-rays and neutrinos. Electrons and
positrons can later lose energy on bremsstrahlung, inverse Compton
scattering and synchrotron processes.

We compute the spectra of $\gamma$-rays from decay of neutral pions produced
in the interaction of such a proton beam with a cloud which enters the jet.
During propagation through the cloud, the e-p beam Lorentz factor $\gamma_b$
changes as a result of collisionless energy losses according to 

\begin{eqnarray} \gamma_b(x) = \gamma_{b,0}\times 2^{-x/L_p}, 
\end{eqnarray}

\noindent 
where $x$ is the propagation distance in the cloud,
$\gamma_{b,0}$ is the initial Lorentz factor of e-p beam.

The differential $\gamma$-ray spectrum from interaction of e-p beam with a 
single cloud can be obtained from
\begin{eqnarray}
N_\gamma\equiv {{dN_\gamma}\over{dE_\gamma dS dt}} =  \int_0^{L} 
\int_{E_{p,min}}^{E_{p,max}}  {{dN_p}\over{dE_p dV}} 
{{dN_\gamma(E_p)}\over{dE_\gamma dt}} dE_p dx
\end{eqnarray}
\noindent
where $dN_p/dE_p dV = A E_p^{-\alpha}$ is the differential density of
relativistic protons (per energy and volume) which is assumed to be a power
law type with the index $\alpha$ and normalization $A$, 
$dN_\gamma(E_p)/dE_\gamma dt$ is the 
differential $\gamma$-ray photon spectrum produced per unit time by 
monoenergetic protons with energy $E_p = m_p \gamma_b$,
$L$ is the maximum propagation distance in the cloud, and $m_p$ is the
proton rest mass. In computations of these spectra the scaling model has been 
used (Stephens \& Badhwar 1981) which gives a good approximation for the p-p 
cross section in the proton energy range considered.

The computations of the $\gamma$-ray spectra have been done for different
initial spectra of proton beam defined by $\alpha$ and $A$. In Fig.~2 we
show such $\gamma$-ray spectra, multiplied by the square of photon energy
and divided by the propagation distance $L$ in the cloud, for $\alpha = 2.$
(Figs. 2a,b,c) and $\alpha = 2.5$ (Figs.~2d,e,f), for different
normalizations $A = 10^4, 10^5$, and $10^6$. Different curves show the
$\gamma$-ray spectra for different propagation distances of the proton beam
in the cloud, and the dotted curve shows the $\gamma$-ray spectrum in the
absence of collisionless losses.  The beam power for the applied proton
spectrum is comparable to the expected jet power in AGN. For example, if
the jet radius is $S = 10^{16}$ cm $^{-2}$, and $\alpha =2$ and $A = 10^5$,
then the proton beam power is $P_b = c S A\int_{E_{th}}^{E_{p,max}} E_p^{-2}
E_p dE_p \approx 10^{46}$ erg s$^{-1}$, for $E_{p,max} = 10^4$ GeV.  Since
the parameters of the cloud determine the propagation distance of the proton
beam, the $\gamma$-ray spectra are computed for the cloud densities $n_c =
10^{12}$ cm$^{-3}$, and $10^{13}$ cm$^{-3}$, and temperatures $T_c =10^4$ K,
and $10^3$ K. Such cloud parameters are expected in the broad line regions
of AGNs (Dar \& Laor 1997). For the range of investigated parameters, the
$\gamma$-ray spectra show a characteristic break located between $10 - 100$
GeV.

\section{Confrontation of the Proton Beam Model with Observations of Mrk 421}

Mrk 421 shows strong flares observed simultaneously in TeV and X-ray energy
range on time scales from days up to 15 min (e.g. Buckley et al. 1996,
Gaidos et al. 1996). The spectrum observed by the Whipple Observatory during
a strong flare on May 7 1996 has the spectral index $2.56\pm 0.07\pm 0.1$
between 0.3 - 10 TeV (McEnery et al. 1997). The spectrum of Mrk 421 observed
during low TeV state by the EGRET telescope between 100 MeV and a few GeV
has the spectral index $1.71\pm 0.15$ (Lin et al. 1994). This spectrum does
not change significantly between the low and high TeV states (Buckley et al.
1996). Similar flaring behavior is also observed in the case of Mrk 501 in
the TeV energy range (see the review by Protheroe et al. 1997), and in
x-rays (Catanese et al. 1997b, Pian et al. 1998). Below we discuss the
expected radiation signatures from the proton - cloud interaction model in
the context of these observations.

\subsection{Gamma-rays from Decay of $\pi^o$}

The existence of a break in the $\gamma$-ray spectra of two BL Lacs (Mrk 421
and Mrk 501) observed in TeV energy range is at present very clear. The
relativistic proton beam model with a single power law spectrum 
is able to explain this feature quite naturally.

In order to find out if the model is able to explain the shape of the
$\gamma$-ray spectrum observed from Mrk 421 in the TeV and GeV energy range,
we compare in Fig.~3 the results of observations with the expected
emission from the beam-cloud collision model. We assume that the strong
variability of the TeV emission is caused by the proton beam propagating
through regions of cloud(s) with different thickness $L$. 

The much lower variability of the MeV - GeV $\gamma$-ray flux is generally
consistent with the expectations of proton beam model. The computed spectra
also show a break between the GeV and TeV energy range, consistent with the
observations of Mrk 421 and Mrk 501. This break is caused by the
collisionless losses of the e-p beam. However if the computed spectrum is
normalized to the spectrum observed in the TeV energies, then the predicted
GeV spectrum is significantly flatter, being inconsistent with the spectral
index in the GeV energies observed by EGRET (Lin et al. 1994, Buckley et al. 
1996). Therefore the $\gamma$-rays produced from decay of neutral pions
cannot by themselves be used to explain the weakly variable GeV emission
observed from Mrk 421 unless the spectrum of protons in the beam 
show strong steepening below $\sim 10$ GeV.  
Contributions from another source of $\gamma$-rays
to the GeV energy range is needed at least at energies below $\sim 1$ GeV.
In the next subsection we discuss  contributions to the emitted radiation
from secondary $e^\pm$ pairs.

\subsection{Photons Produced by Secondary $e^\pm$ Pairs from the Decay of 
$\pi^\pm$}

If jet-cloud model is valid for the $\gamma$-ray production in Mrk 421, 
then the secondary $e^\pm$ pairs from decay of charged pions should have a
spectrum very similar to the $\gamma$-ray spectrum from $\pi^0$ decay, but
shifted to lower energies by a factor of two. Let us assume, following
Dar \& Laor (1997), that these secondary $e^\pm$ pairs are responsible for
the production of synchrotron emission during the outburst in Mrk 421. In fact
the equilibrium spectra of secondary $e^\pm$ pairs are consistent with the 
synchrotron spectrum observed from Mrk 421 in a high state. The break in the 
spectrum of secondary pairs at $E_{\pm,b}$ should correspond to the break in 
the synchrotron spectrum which is observed in Mrk 421 at 
$\sim 1.65$ keV (Takahashi et al. 1996). 
The relation between these breaks allows us to estimate the strength of 
the magnetic field in the region of production of synchrotron photons from 

\begin{eqnarray}
B\approx 2B_{cr}\epsilon_b m_e E_{e^\pm,b}^{-2}, 
\end{eqnarray}

\noindent where $B_{cr} = 4.414\times 10^{13}$ Gs, and $m_e$ is the electron
rest mass. For the value $E_{e^\pm,b}\sim 100$ GeV (see Fig.~3), the
magnetic field in the emitting region should be equal to $B\approx 7.1$ G.
The characteristic cooling time of secondary pairs with energies mensioned
above in such a magnetic field
is of the order of $\sim 70$ s which is consistent with the time scale of
outbursts in Mrk 421.

The x-ray and $\gamma$-ray flares have similar powers in Mrk 421. Therefore
the secondary pairs must move almost ballistically through the cloud, e.g.
by following the highly ordered magnetic field lines along the jet axis. 
The pairs move through the dense cloud during the characteristic time
$t_c\approx r_c/c$ which is comparable to their 
characteristic synchrotron cooling time estimated below.
Therefore, the observed synchrotron
emission is produced close to the cloud and should be synchronized with
the TeV $\gamma$-ray emission.

The comparison of synchrotron and bremsstrahlung cooling times of secondary
$e^\pm$ pairs shows that  
bremss\-tra\-hlung dominates for $e^\pm$ pairs with Lorentz factors

\begin{eqnarray}
\gamma < \gamma_{bs}\approx 4.4\times 10^{-7} n_HB^{-2},
\end{eqnarray}

\noindent where $n_H$ is the density of the background matter. For the value
of the magnetic field estimated above, and considered cloud
density $n_H = 10^{12}$ cm$^{-3}$, we get $\gamma_{bs}\approx 9\times
10^{4}$. However the pairs travel through the cloud during the time $t_c$,
which is shorter than their cooling time via bremsstrahlung:
$\tau_b\approx 1.4\times 10^{15} n_H^{-1}$ s. Hence we conclude that 
secondary $e^\pm$ pairs, with Lorentz
factors $\gamma < \gamma_{br}$, have no time to cool by a bremsstrahlung
process in the cloud.  This process cannot, therefore, contribute essentially
to the Mrk 421 spectrum in the MeV - GeV energy range.

The contribution to the $\gamma$-ray spectrum from inverse Compton
scattering (ICS) of soft photons by secondary pairs is very uncertain
because of the lack of precise information on the soft photon geometry and
density in the emission region.  The observations of Mrk 421 by the EGRET
and ASCA telescopes allow us to estimate the ratio of the power emitted in MeV
- GeV $\gamma$-rays and x-rays, which is close to 0.5 (see von Montigny et
al. 1995, Takahashi et al. 1996).  Therefore if ICS of secondary pairs
contributes significantly to the EGRET energy range, the energy densities of
the magnetic field and the low-energy radiation in the region of the propagation
of the secondary pairs has to be in a similar ratio, provided that the
scattering occurs in the Thomson regime.  The energy density of the magnetic
field in the emission region of Mrk 421 (see above) requires that the energy
density of soft photons should be $\rho_{ph}\approx 8\times 10^{11}$ eV
cm$^{-3}$. 

Three different sources of soft photons can provide, in principle, a target
for these secondary pairs, i.e. soft synchrotron photons, thermal photons
coming directly from an accretion disk, or thermal bremsstrahlung photons 
produced in the cloud. Two first possibilities can be
excluded based on following arguments. Let us assume that the stable flux of
$\gamma$-rays in the 100 MeV - 1 GeV energy range is produced by ICS in the
Thomson regime. The scattering in the Klein-Nishina regime reduces the
efficiency of the process and produces too steep a spectrum. The part of the
spectrum of secondary pairs below $\sim 10$ GeV does not vary strongly (see
Figs.~3). However, these pairs can produce 100 MeV - 1 GeV $\gamma$-rays by
scattering synchrotron x-ray photons, which are highly variable. Therefore
the flux of $\gamma$-rays in the EGRET energy range should vary
significantly between the low and high TeV state, which is in contradiction
with the observations (Lin et al. 1994, Buckley et al. 1996).  The flux of
secondary pairs with energies above $\sim 10$ GeV is highly variable and
cannot produce EGRET $\gamma$-rays by scattering weakly variable optical -
UV photons of synchrotron origin, or produced in the low temperature
optically thick accretion disk. 

The density of thermal bremsstrahlung photons produced in the cloud can be
estimated from
\begin{eqnarray}
\rho_{br}\approx 3\times 10^{-26} T_c^{0.5} n_c^2 r_c {\rm ~eV cm}^{-3}.
\end{eqnarray}
\noindent
However these photons have characteristic energies $\epsilon_{br}\sim 1$ eV. 
Hence they require electrons with Lorentz factors $\sim 3\times 10^4$ in order
to produce 1 GeV $\gamma$-rays in ICS process. Secondary pairs with these
energies show already high level of variability between low and high states
of Mrk 421, and therefore should produce strongly variable GeV emission which
is in contrary to observations.

The only possibility left is that the
secondary pairs with energies below $\sim 10$ GeV scatter non-variable
radiation with characteristic energies of $\epsilon_X = 0.1 - 1$ keV of
another origin (e.g., produced in the high temperature disk or the disk
corona). However, this radiation field has to be transparent to the observed
$\gamma$-rays. The optical depth for the GeV $\gamma$-rays is less than one
if the characteristic dimension of this X-ray region is 

\begin{eqnarray} r_X < {{\epsilon_X}\over{\rho_{ph} \sigma_{\gamma\gamma}}}
\approx 5\times 10^{15}{\rm cm}, \end{eqnarray} \noindent where
$\sigma_{\gamma\gamma}$ is the maximum value of the photon-photon pair
production cross section and applying $\epsilon_X = 1$ keV. This radiation
might eventually originate in the inner part of the accretion disk or the
disk corona or might be produced by the low density but high temperature
plazma surrounding small and dense clouds which are considered here as a
target for relativistic proton beam. However the above constraint on the
dimension of this region rather exclude the last possibility.

\section{Conclusion}

If the relativistic electron-proton beam collides with a dense, compact cloud, the
energy losses of the jet via the excitation of plasma waves (Rose et al
1984) becomes important for beams with energies below $\sim 300$ GeV
(Fig.~1). This process causes the break in the spectrum of $\gamma$-rays
from the decay of neutral pions that are produced in inelastic collisions of
protons with matter. Such a model can naturally explain the low and high
states of TeV $\gamma$-ray emission in BL Lac type blazars, provided that
the proton beam interacts with different column densities of matter. Moreover,
because of the collisionless energy losses of the proton beam, the
$\gamma$-ray spectrum predicted in the 100 MeV - 1 GeV energy range does not
vary strongly.  This features fits nicely to the general behavior of the
$\gamma$-ray emission from Mrk 421 between low and high states. However the
$\gamma$-ray spectra computed in the 100 MeV - 1 GeV energy range for
reasonable sets of model parameters are too flat, showing significant
deficiency in comparison to the Mrk 421 spectrum observed by EGRET
(Fig.~3) provided that the spectrum of relativistic protons in the beam is well 
described by a single power law. Secondary $e^\pm$ pairs are also produced in hadronic
collisions through decay of charged pions, with the spectra similar to the
$\gamma$-ray spectra but shifted to lower energies. 
The higher energy, variable part of the spectrum of secondary
pairs (with energies above $\sim 10$ GeV) can be responsible for the
simultaneous x-ray flares.

Such a picture could work provided that the clouds are not too hot.
Otherwise the TeV $\gamma$-rays
are absorbed by the thermal bremsstrahlung photons produced in the cloud
(see computations of the $\gamma$-ray mean free paths on Fig.~3 in Beall
et al. 1987). The clouds also cannot be too dense. By reversing Eq.~(9),
we see that if the cloud density is
\begin{eqnarray}
n_c >  3\times 10^{25} \rho_{ph} T_c^{-0.5} \lambda^{-1},
\end{eqnarray}
\noindent
where $\lambda = n_c r_c$ is the column density of matter traversed by the
proton beam in the cloud,
then the secondary electrons are not able to lose energy efficiently
by the synchrotron process because of dominant role of ICS losses on thermal
bremsstrahlung radiation produced in the cloud.

The relation of the break in the spectrum of secondary pairs to the break in
the x-ray spectrum observed in Mrk 421 allows a derivation of the strength
of the magnetic field in the emission region equal to $\sim 7.1$ G (Eq.~3). 
These pairs cannot contribute to the EGRET energy range: by production of
$\gamma$-rays in a bremsstrahlung process; or by inverse-Compton scattering
of soft photons of synchrotron origin, or coming from the optically thick,
low temperature accretion disk, or bremsstrahlung photons produced in the
cloud.  Only the scattering of non-variable, soft x-ray photons ($0.1 - 1$
keV) by secondary pairs with energies below $\sim 10$ GeV could explain the
non-variable $\gamma$-ray emission from Mrk 421 in the EGRET energy range.
These photons could be produced in the inner part of the accretion disk.

\section*{acknowledgments} 

WB would like to thank the Institute for
Computational Sciences and Informatics at George Mason University (Fairfax,
VA) for hospitality during his visit.  

JHB gratefully acknowledges the assistance of an AAS Small Research Grant and
a grant from the Newstead Foundation which was helpful in support of this
research.

\eject

\eject

   \begin{figure}[t]
      \vspace{17truecm}
\includegraphics{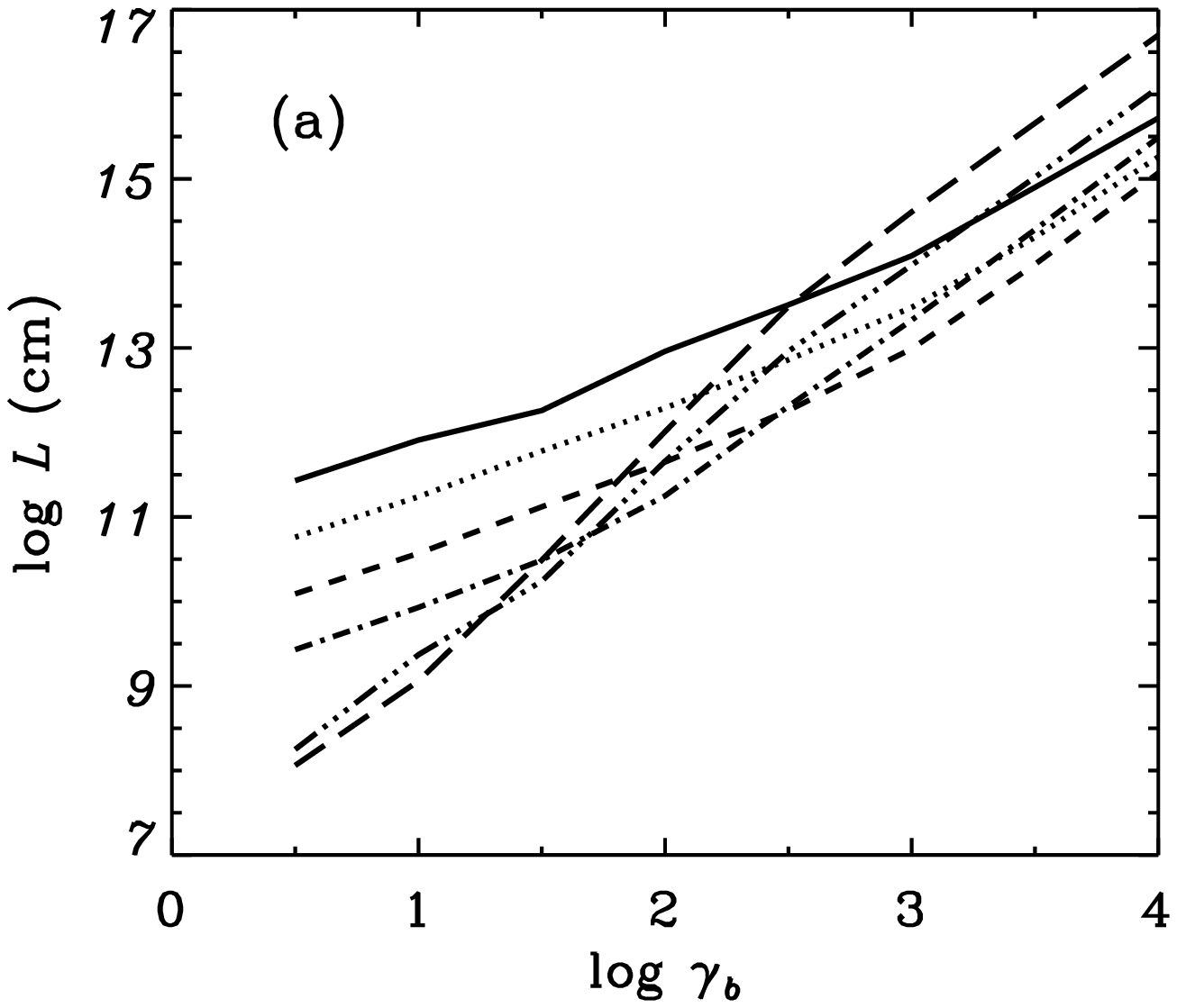}
      \caption[]{Characteristic propagation distances $l$ for the 
electron-proton beam as a function of its Lorentz factor. Specific curves
show the results for beam densities $n_b = 1$ cm$^{-3}$ (full curve), 
10 cm$^{-3}$ (dotted curve), $10^2$ cm$^{-3}$ (dashed curve), $10^3$ cm$^{-3}$
(dot-dashed curve), $10^4$ cm$^{-3}$ (three dot-dashed curve), $10^5$ cm$^{-3}$
(long-dashed curve). Fig.~1 shows the results for temperature of the cloud  
$T = 10^4$ K, and its density $n_c = 10^{12}$ cm$^{-3}$. 
}
	 \label{fig1}
    \end{figure}

\newpage

   \begin{figure*}[t]
      \vspace{17truecm}
\includegraphics{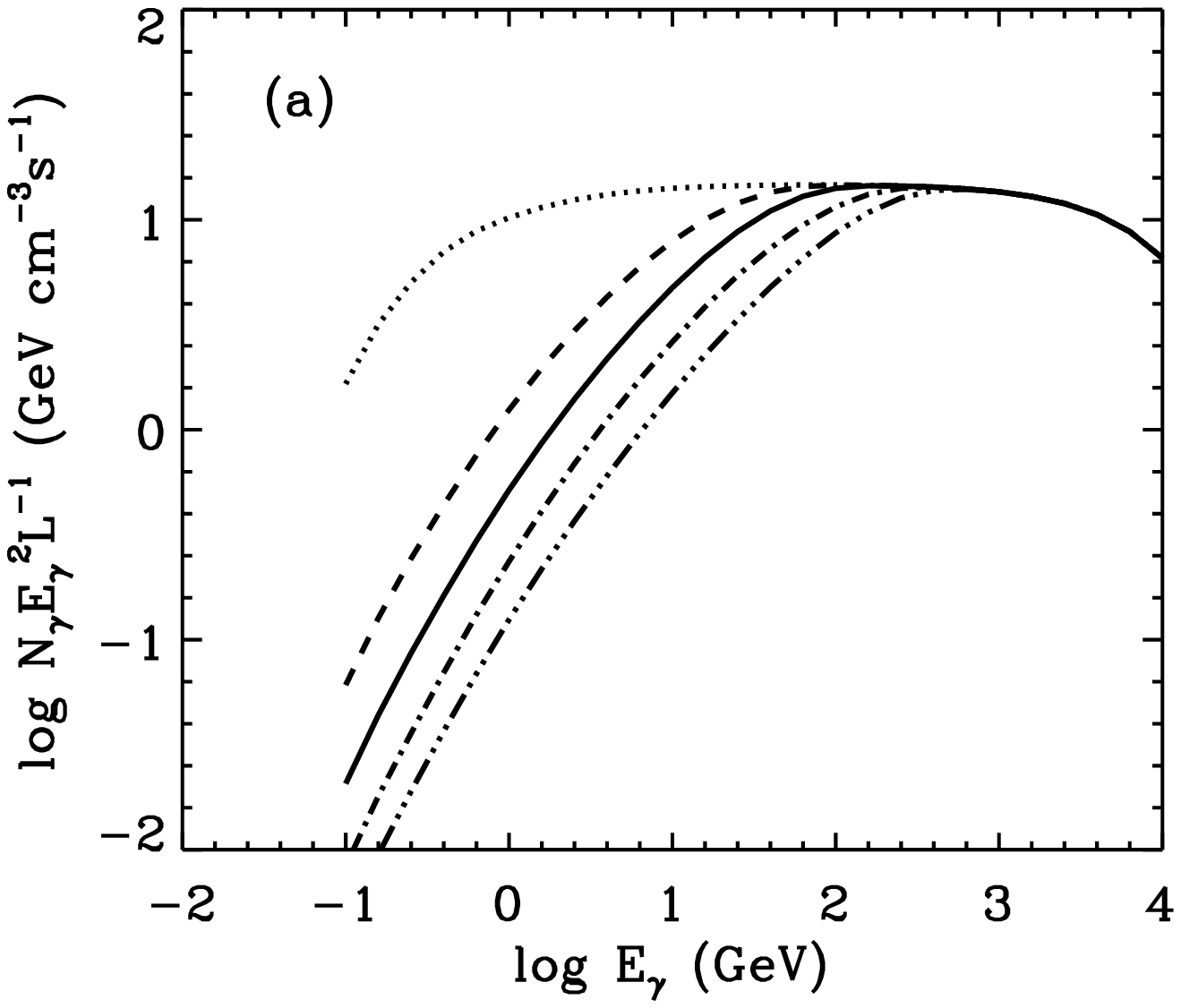}
\includegraphics{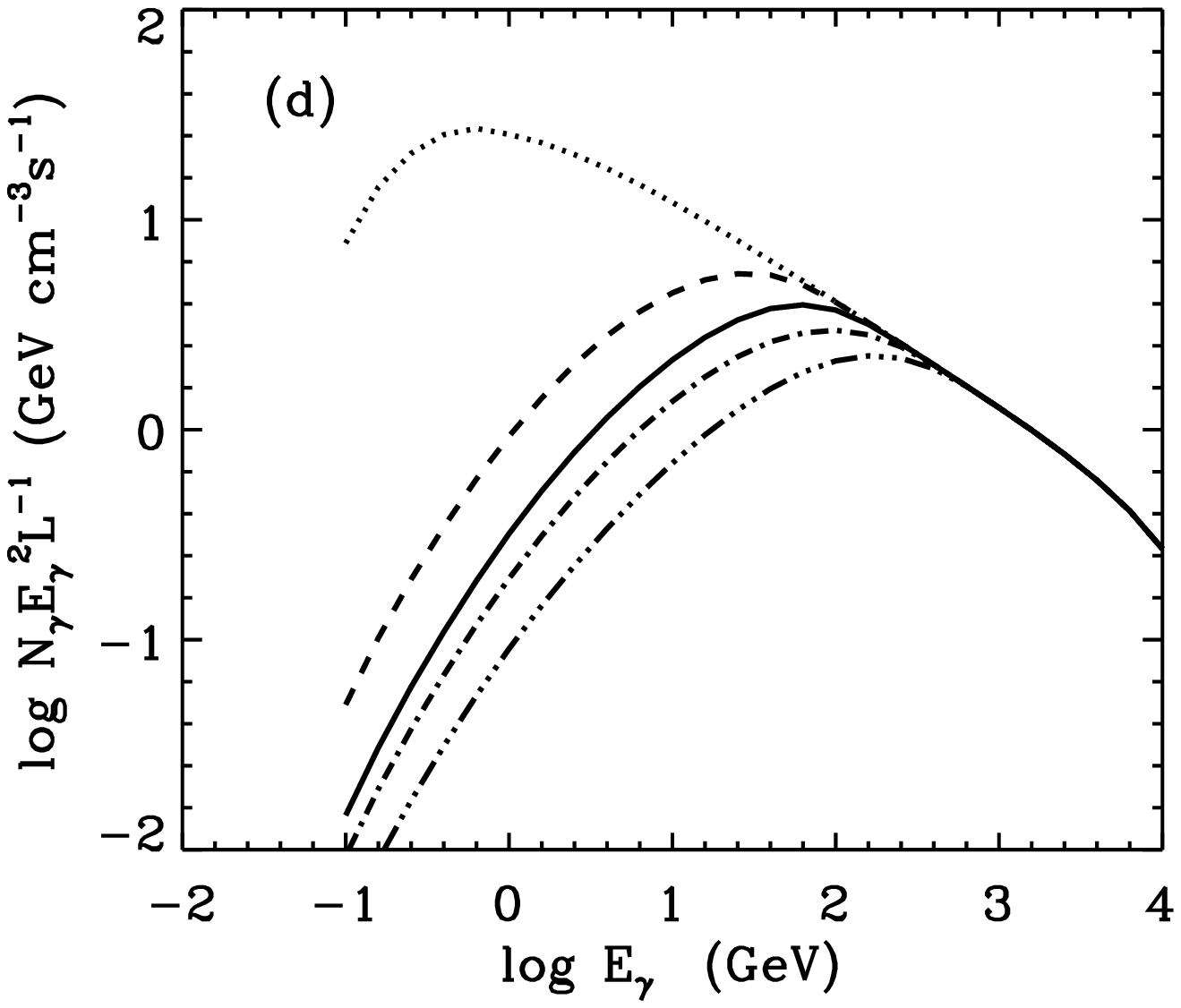}
\includegraphics{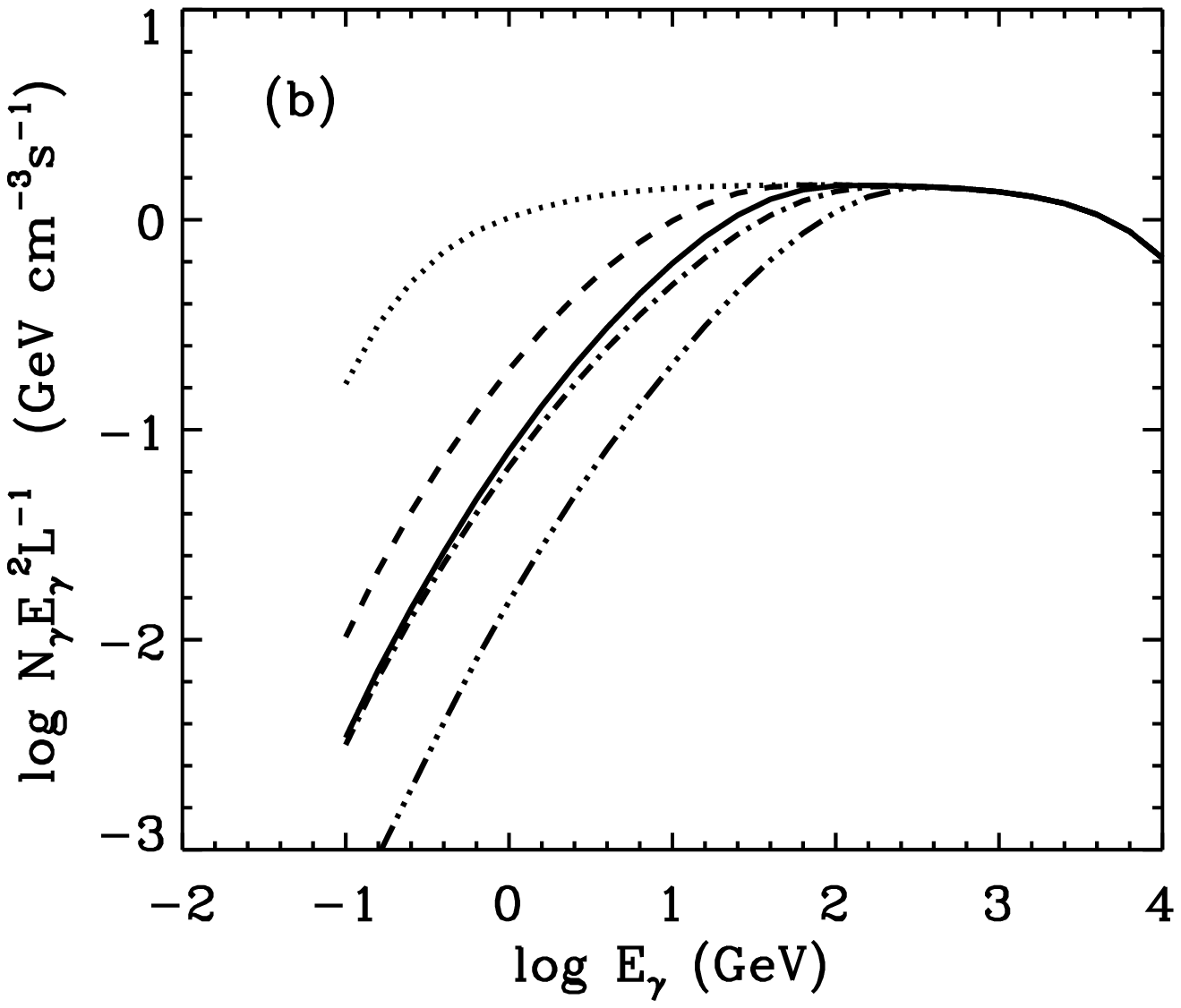}
\includegraphics{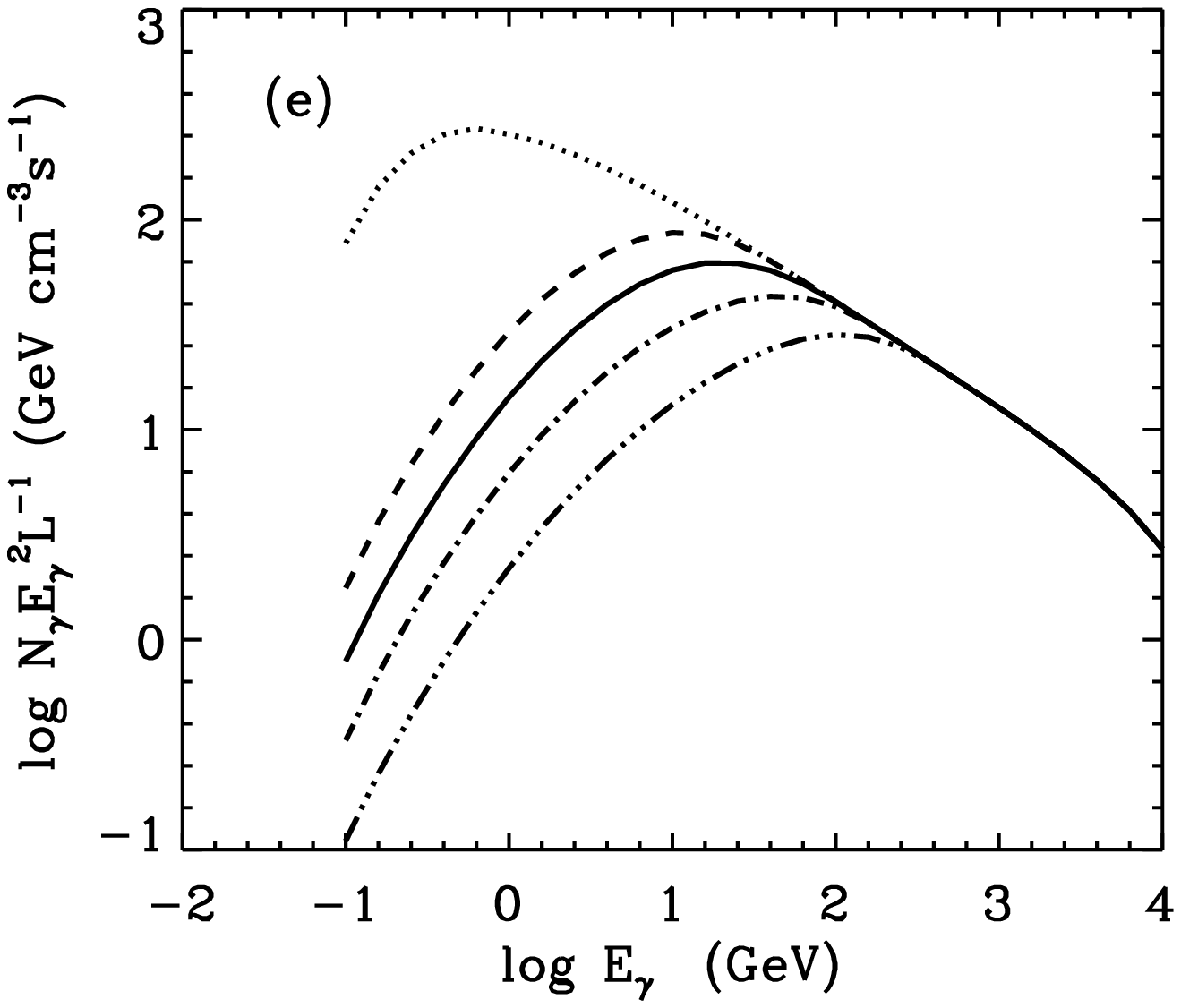}
\includegraphics{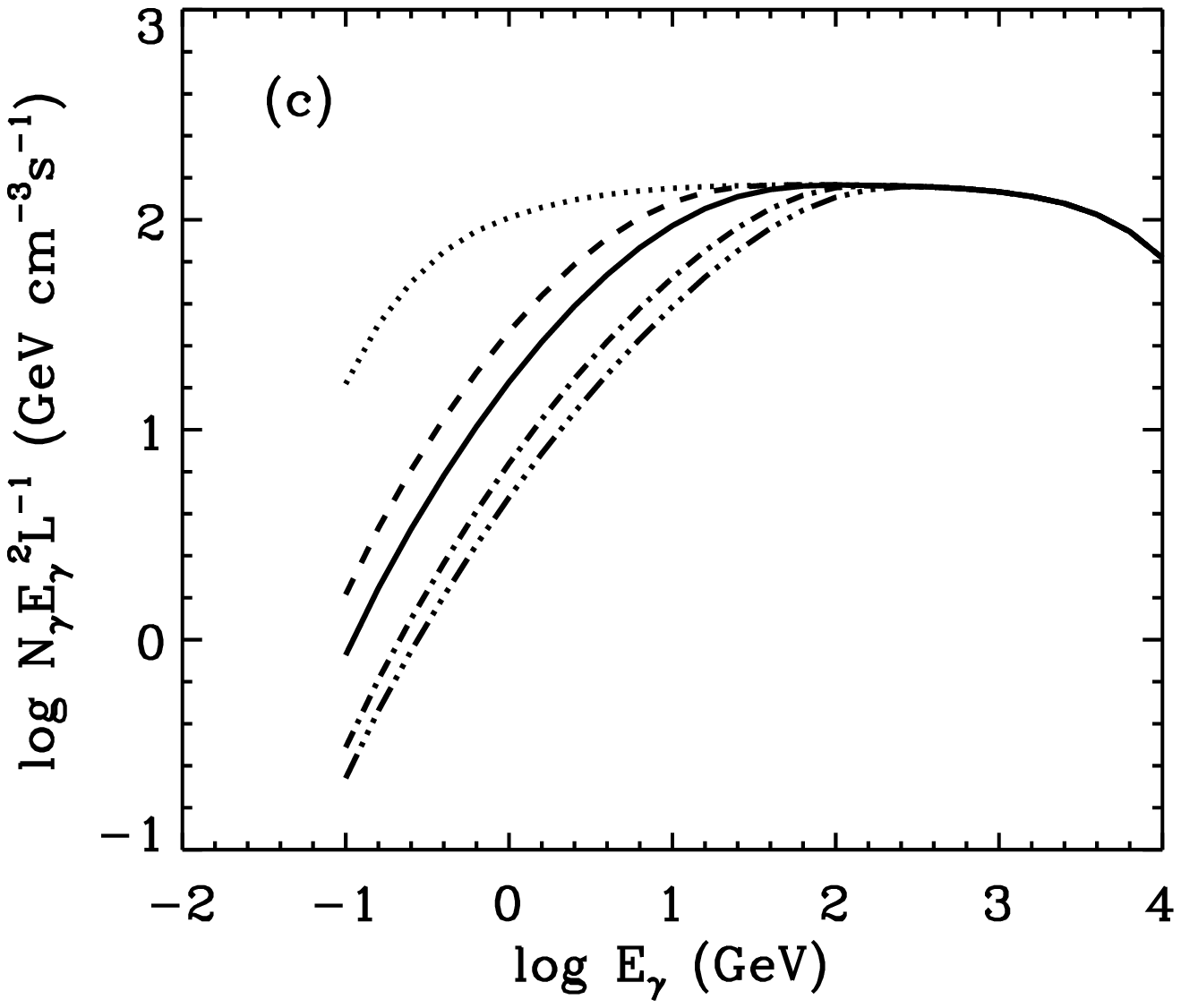}
\includegraphics{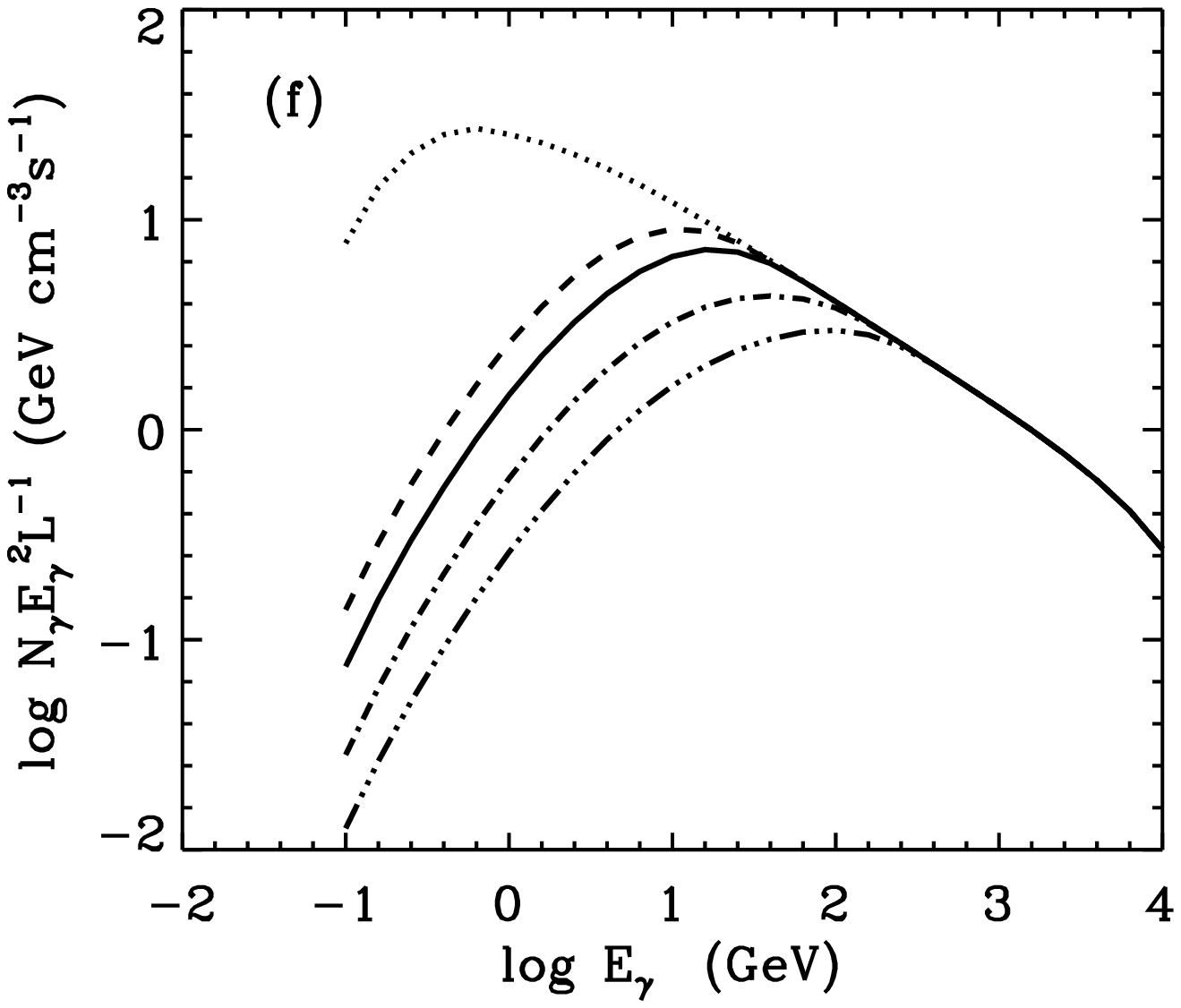}
      \caption[]{(a) The $\gamma$-ray spectra produced during propagation of
relativistic electron-proton beam in the cloud with inclusion of
collisionless losses. These results are shown for the power law spectrum of
protons with the spectral index $\alpha=2$ (see (a), (b), and (c)) and
$\alpha = 2.5$ (see (d), (e), (f)) and normalization $A = 10^4$ (b), $10^5$
(a,), and $10^6$ (figures c,d,e,f). The proton beam propagates in the cloud
with the density $n_c = 10^{12}$ cm$^{-3}$ (a,b,d,f), and $n_c = 10^{13}$
cm$^{-3}$ (c,e), and temperature $T_c = 10^3$ K (f), and $T_c = 10^4$ K in
other figures . The specific curves correspond to different propagation
distances in the cloud $L = 10^{12}$ cm (dashed curve), $3\times 10^{12}$ cm
(full curve), $10^{13}$ cm (dot-dashed curve), and $3\times 10^{13}$ cm,
accept (c) and (e) for which the propagation distances are an order of
magnitude lower. The dotted curve shows the $\gamma$-ray spectrum in the
case when collisionless losses are not included. }
	 \label{fig2}
    \end{figure*}

\newpage 

   \begin{figure}[t]
      \vspace{17truecm}
\includegraphics{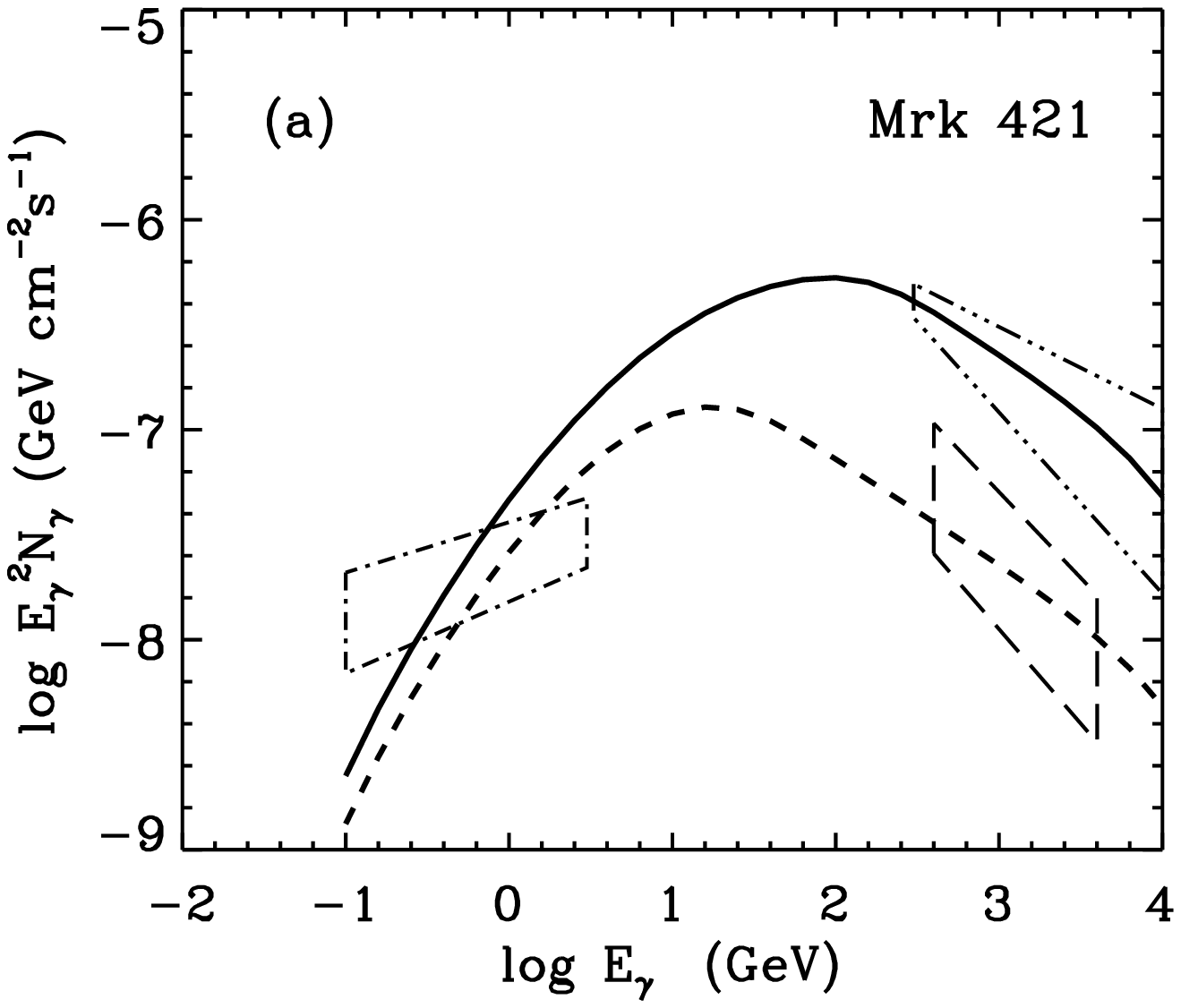}
\includegraphics{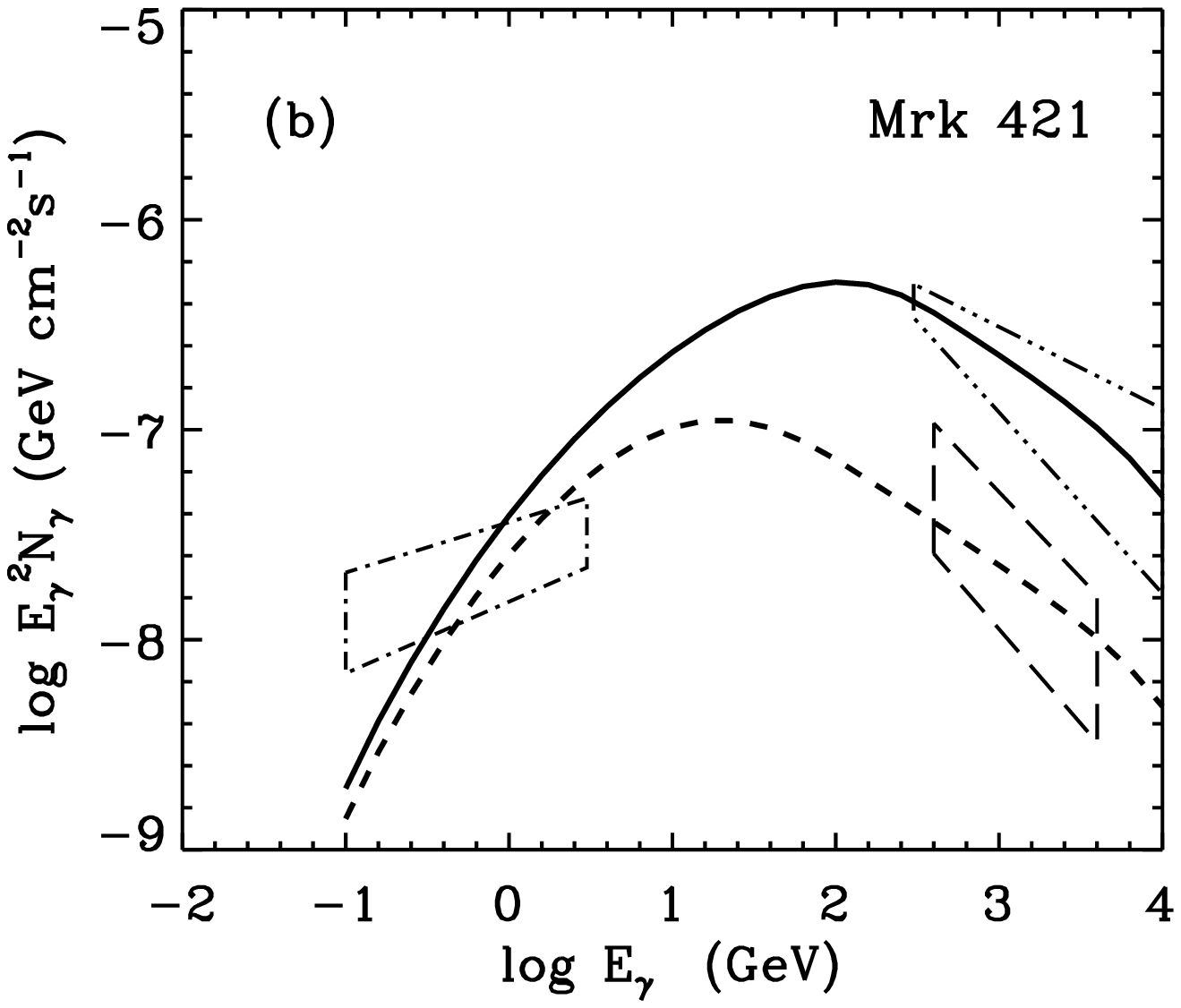}
\includegraphics{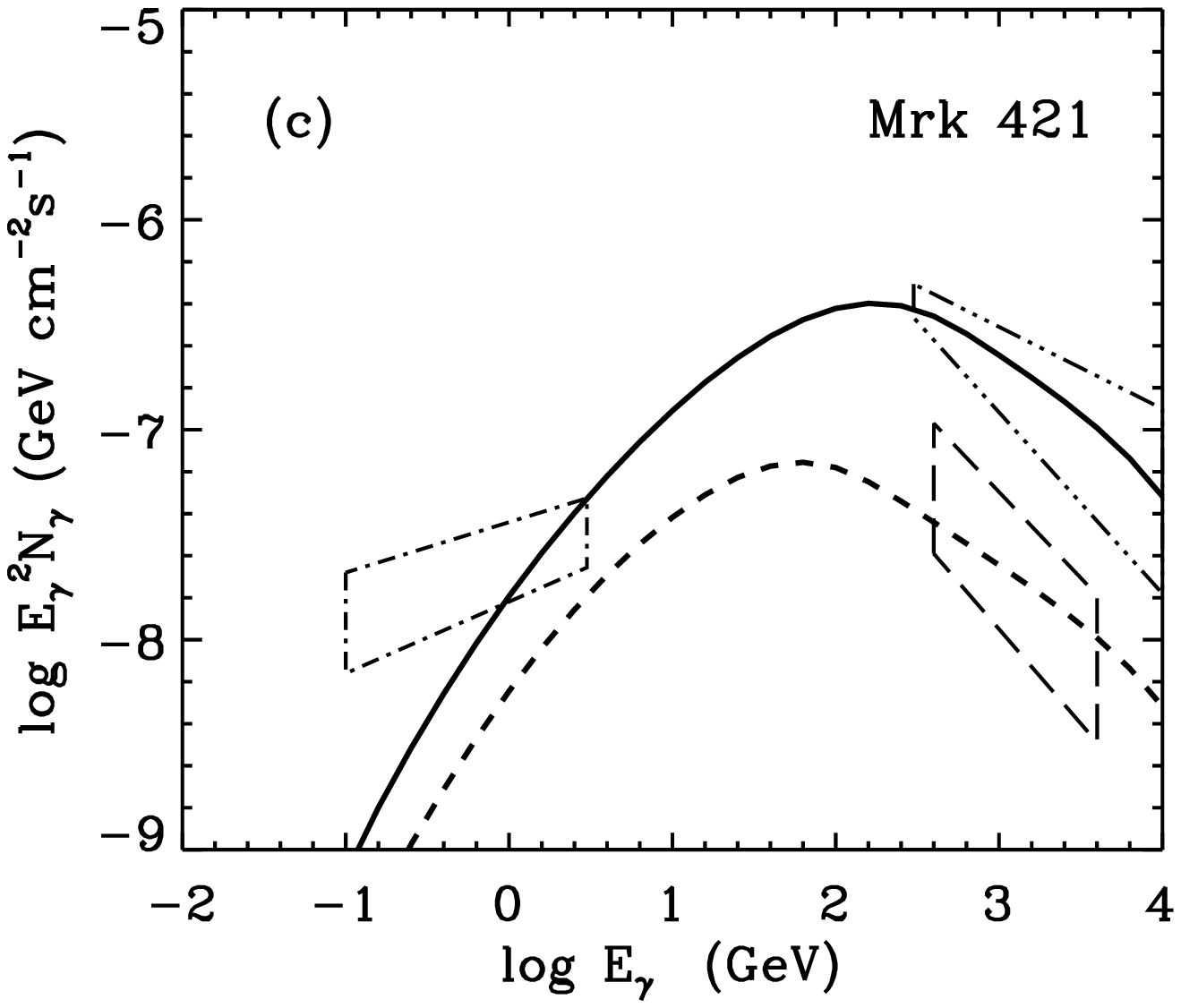}
      \caption[]{The differential $\gamma$-ray spectra from decay of neutral
pions produced by proton beam with the power law spectrum and index 2.5 are
compared with the observations of Mrk 421 by: the EGRET telescope
(dot-dashed box, Lin et al.~1994), the Whipple Observatory in a low state
(dotted box, Mohanty et al.~1993), and in a high state (dot-dot-dot-dashed
box, McEnery et al.~1997). Two $\gamma$-ray spectra correspond to different
propagation distances in the cloud equal to $L = 3\times 10^{13}$ cm (full
curve) and $L = 3\times 10^{12}$ cm (dashed curve) in Figs.~3a,c and for an
order of magnifute lower propagation distances in Fig.~3b. The parameters of
the proton beam and the cloud in Figs. 3a,b,c are these same as in Figs.
2f,e,d, respectively. }
	 \label{fig3}
    \end{figure}

\end{document}